\documentclass[pra, nofootinbib, twocolumn, superscriptaddress]{revtex4-2}
\usepackage{amsmath}
\usepackage{amssymb}
\usepackage{graphics}
\usepackage{epsfig}
\usepackage{verbatim}
\usepackage{textcomp}
\usepackage[colorlinks=true,urlcolor=blue,linkcolor=blue,citecolor=blue]{hyperref}
\usepackage{tabularx}
\usepackage{xcolor}

\begin{document}
	
	\title{Majorana's approach to nonadiabatic transitions validates the adiabatic-impulse approximation}
	
	
	\author{P.~O.~Kofman}
	\affiliation{B.~Verkin Institute for Low Temperature Physics and Engineering, Kharkiv 61103, Ukraine}
	\affiliation{V.~N.~Karazin Kharkiv National University, Kharkiv 61022, Ukraine}
	\affiliation{Theoretical Quantum Physics Laboratory, Cluster for Pioneering Research, RIKEN, Wakoshi, Saitama 351-0198, Japan}
	
	\author{O.~V.~Ivakhnenko} 
	\affiliation{B.~Verkin Institute for Low Temperature Physics and Engineering, Kharkiv 61103, Ukraine}
	\affiliation{Theoretical Quantum Physics Laboratory, Cluster for Pioneering Research, RIKEN, Wakoshi, Saitama 351-0198, Japan}	
	
	\author{S.~N.~Shevchenko}
	\email{sshevchenko@ilt.kharkiv.ua}
	\affiliation{B.~Verkin Institute for Low Temperature Physics and Engineering, Kharkiv 61103, Ukraine}
	\affiliation{V.~N.~Karazin Kharkiv National University, Kharkiv 61022, Ukraine}

	\author{Franco~Nori}
	\affiliation{Theoretical Quantum Physics Laboratory, Cluster for Pioneering Research, RIKEN, Wakoshi, Saitama 351-0198, Japan}
	\affiliation{Quantum Computing Center, RIKEN, Wakoshi, Saitama 351-0198, Japan}
	\affiliation{Department of Physics, The University of Michigan, Ann Arbor, MI 48109-1040, USA}	
	
	\begin{abstract}
		The approach by Ettore Majorana \cite{Majorana1932} for non-adiabatic transitions between two quasi-crossing levels is revisited. We rederive the transition probability, known as the Landau-Zener-St\"{u}ckelberg-Majorana formula, and introduce Majorana's approach to modern readers. This result, typically referred as the Landau-Zener formula, was published by Majorana before Landau, Zener, St\"{u}ckelberg. Moreover, we obtain the full wave function, including its phase, which is important nowadays for quantum control and quantum information. The asymptotic wave function correctly
		describes dynamics far from the avoided-level crossing, while it has limited accuracy in that region.
	\end{abstract}
	
	\date{\today }
	\maketitle
	\section{Introduction}
	
	\label{Sec:Introduction} A few years after the discovery of the Schr\"{o}dinger equation, it was solved for the problem of transitions between two
	energy levels~\cite{Majorana1932, Landau1932a, Landau1932b, Zener1932, Stueckelberg1932}, and the solution is now known as the Landau-Zener-St\"{u}ckelberg-Majorana (LZSM) formula. We return to this non-trivial problem in
	the view of modern interest in quantum control and quantum information \cite{Nakamura2011,
		Shevchenko2019, Ivakhnenko2022}.
	
	The contributions of these four authors have been discussed in the recent literature with detailed derivations. Here we give a few references: for Landau's approach, see
	Ref.~\cite{LL}; for Zener's approach see Ref.~\cite{Shevchenko2010} and
	references therein; for St\"{u}ckelberg's approach see Refs.~\cite{Child1974, Child1974a}; and for Majorana's approach see Ref.~\cite%
	{Wilczek2014}. All the four approaches give exactly the same LZSM formula
	for the transition probability \cite{DiGiacomo2005, Ivakhnenko2022}. But can we derive the \textit{full}, including phase, wave function with these
	approaches? The answer is negative for Landau's approach and for St%
	\"{u}ckelberg's approach \cite{Child1974}. It is well known, that Zener's
	approach does give the full wave function \cite{Vitanov1996}. However, to the best of our knowledge, this question has not been fully addressed in
	literature for Majorana's approach, cf. Refs.~\cite{Wilczek2014,Rodionov2016,Dogra2020}. In particular, recently it was pointed out that
	Majorana's contribution to the problem has been underestimated \cite%
	{DiGiacomo2005, Wilczek2014}.
	
	The author of Ref.~\cite{Wilczek2014} discusses from a modern perspective
	several of Majorana's works related to condensed matter physics, and among
	these, the paper of interest here: Ref.~\cite{Majorana1932}. Addressing this, Ref.~\cite{Wilczek2014} rederives the LZSM formula; however does not find the full wave
	function, including the phase, after the passage of the avoided-level
	crossing. The full asymptotic wave function was found in Ref.~\cite%
	{Rodionov2016} following Majorana's approach. There,
	the authors \cite{Rodionov2016} studied both the direct and inverse transition of the
	avoided-level crossing, found the asymptotic wave functions, and introduced the transfer-matrix (or adiabatic-impulse) method to describe the evolution.
	
	\begin{figure}[t]
		\centering{\includegraphics[width=1\columnwidth]{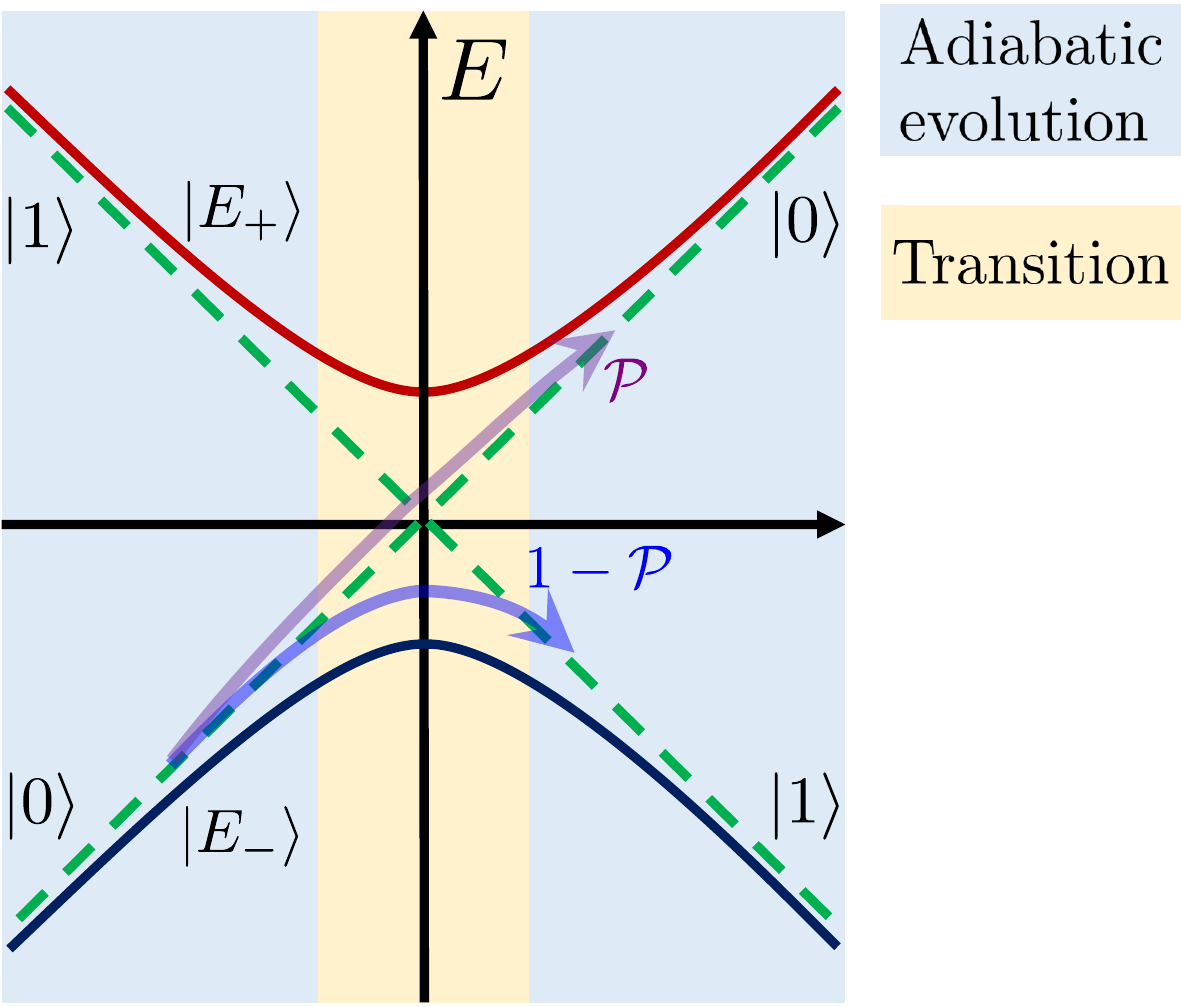}}
		\caption{The two-level system and the adiabatic-impulse model. The
			approximation consists in the assumption that the evolution is described by the alternation of stages. During the adiabatic stages the system follows either a ground state $E_{-}$\ or an excited state $E_{+}$. The impulse-type transition (with the probability $\mathcal{P}$) is assumed to happen in the point of the minimal energy-level
			splitting. In this work, the adiabatic-impulse model is justified within
			Majorana's approximation.}
		\label{Fig:Adiabatic Impulse Model}
	\end{figure}
	
	However, several important questions are left unanswered. For one thing,
	\textquotedblleft \textit{it is interesting to contemplate an ambitious
		program, where one might infer the time-dependence of states from the
		time-dependent energy levels}\textquotedblright ~\cite{Wilczek2014}. Does Majorana's approach give the correct description of the \textit{dynamics}? And not only the asymptotic behaviour at infinity. Following Refs.~\cite{Majorana1932, Wilczek2014, Rodionov2016}, we explore in detail Majorana's approach. Not only we present how to obtain the asymptotic transition probability at infinity, but
	we also introduce and describe the adiabatic-impulse model and explore the dynamics near the avoided-level crossing.
	
	The energy levels are schematically shown in Fig.~\ref{Fig:Adiabatic Impulse Model}.
	Consider a two-level system with states $\left\vert 0\right\rangle $ and 
	$\left\vert 1\right\rangle $. If starting, say, to the left in a ground
	state (more generally, in a superposition state), what is the transition
	probability to the excited state to the right? What will be the full wave
	function far from the avoided-level crossing to the right? The answer to
	this latter question would help to justify a convenient \textit{adiabatic-impulse model}. This model assumes that
	the transition region is considered  to be infinitely thin, while elsewhere the evolution is
	modelled as adiabatic.
	
	The rest of this work is organised as follows. In Sec.~\ref{DAILT} (with details in Appendix~\ref{AppendixA}) we follow Majorana's methodology and,
	solving the Schr\"{o}dinger equation, obtain the time-dependent wave
	function. The asymptotic behaviour of this, at $t\rightarrow \infty $, is
	further analyzed in Sec.~\ref{PPAAIM} (with details in Appendix~\ref{AppendixB}), resulting in the LZSM transition probability, Stokes phase change, and the convenient adiabatic-impulse model. The solution following Majorana is compared with the one following Zener in Sec.~%
	\ref{CWZA} (with details in Appendix~\ref{AppendixC}). Further
	developments of the Majorana's approach are the time evolution in Sec.~\ref%
	{Dynamics} and using a superposition initial state in Sec.~\ref{AIS}.
	
	Note that the English version of Ref.~\cite{Majorana1932} is available in
	the book \cite{Bassani2006} and also in the second edition in Ref.~\cite%
	{Cifarelli2020}, commented by M.~Inguscio \cite{Inguscio2020}. For
	Majorana's biography and research see Ref.~\cite{Esposito2014} and also the
	book \cite{Esposito2017}.
	
	\section{Direct and inverse Laplace transforms}
	\label{DAILT} In his work \cite{Majorana1932}, Ettore~Majorana studied an
	oriented atomic beam passing a point of a vanishing magnetic field; the
	second half of the paper was devoted to a spin-1/2 particle in a linearly
	time-dependent magnetic field. The author considered the problem about a
	spin orientation in a dynamic magnetic field with components $H_{x}\sim -\Delta $, $H_{y}=0$, and $H_{z}\sim -vt$, where $\Delta $ and $v$
	are constant values.
	
	The time evolution of a linearly driven two-level system is governed by the Schr\"{o}dinger equation: 
	\begin{equation}
		i\hbar \frac{\partial }{\partial t}\left\vert \psi \right\rangle =-\frac{1}{2%
		}\left( \Delta \sigma _{x}+vt\sigma _{z}\right) \left\vert \psi
		\right\rangle ,
	\end{equation}%
	where $\sigma _{i}$ are the Pauli matrices.
	
\begin{figure}[t]
		\includegraphics[width=1\columnwidth]{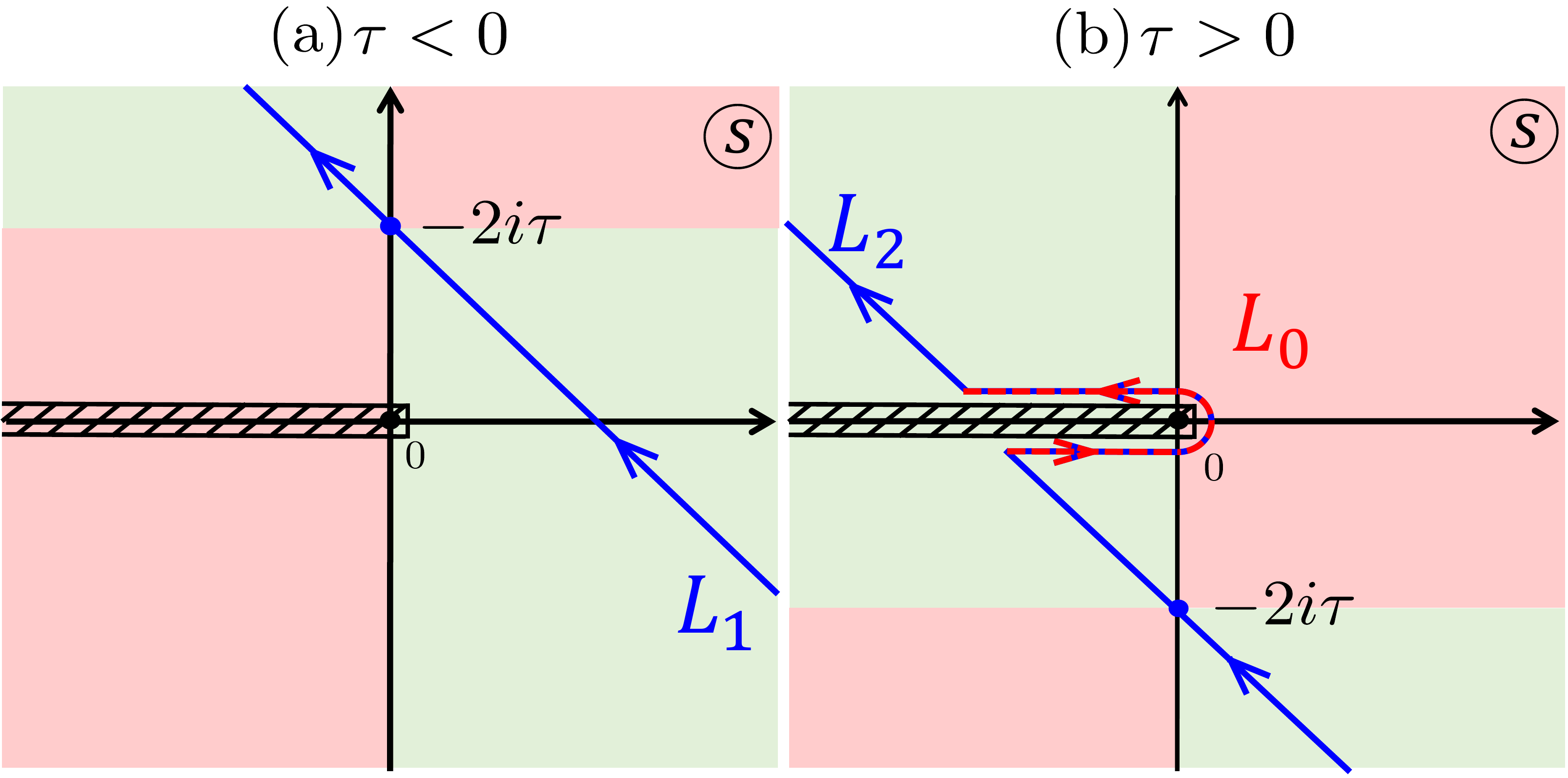}
		\caption{Contours of integration $L_{1,2}$ in Eq.~(\protect\ref{f}) as used
			in Ref.~\protect\cite{Majorana1932}. The regions where the saddle point
			method could be applied are shown in green. The contour $L_{1}$ in~(a) corresponds to $%
			\protect\tau <0$, and $L_{2}$ in~(b) corresponds to $\protect\tau >0$. The
			contour $L_{0}$ is a part of the contour $L_{2}$, which is partly situated in
			the red region. The integration along this contour should be calculated
			separately from the integral calculated within the saddle-point method, both
			of which contribute to the integral in Eq.~(\protect\ref{f}).}
		\label{Fig:Majorana contour}
	\end{figure}

	For the wave function in the form
	
	\begin{equation}
		\left\vert \psi \right\rangle = 
		\begin{pmatrix}
			\alpha \\ 
			\beta%
		\end{pmatrix}%
	\end{equation}
	this can be rewritten as a system of two coupled ordinary differential
	equations. Introducing the dimensionless time $\tau $ and the adiabaticity
	parameter $\delta $ 
	\begin{equation}
		\tau =\sqrt{\frac{v}{2\hbar }}t\text{, \ \ }\delta =\frac{\Delta ^{2}}{4v\hbar },
	\end{equation}
	and making the substitutions 
	\begin{equation}
		\alpha =f\exp\!{\left( \frac{i}{2}\tau ^{2}\right) },\text{ }\beta =g\exp\!{\left( -\frac{i}{2}\tau ^{2}\right) },  \label{MajSubst}
	\end{equation}
	we obtain
	
	\begin{equation}
		\begin{cases}
			\dot{f}\!& =i\sqrt{2\delta }g\exp {(-i\tau ^{2}),} \\ 
			\dot{g}\!& =i\sqrt{2\delta }f\exp {(i\tau ^{2}).}%
		\end{cases}
		\label{initialsystem}
	\end{equation}%
	These can be rewritten for $f$ and $g$ separately 
	\begin{eqnarray}
		\frac{d^{2}f}{d\tau ^{2}}+2i\tau \frac{df}{d\tau }+2\delta f &=&0,
		\label{MajoranaDiffEq1} \\
		\frac{d^{2}g}{d\tau ^{2}}-2i\tau \frac{dg}{d\tau }+2\delta g &=&0.
		\label{MajoranaDiffEq2}
	\end{eqnarray}%
	The substitution (\ref{MajSubst}) is used for obtaining these equations in
	homogeneous form. In this form the Laplace transform simplifies the
	equations. (Note that to obtain Majorana's equations we have to replace $\tau\rightarrow \sqrt{2}\tau_{\mathrm{M}} $ and $\sqrt{\delta}\rightarrow -\sqrt{k}/2$, where $\tau _{\mathrm{M}}$ and $k$ is
	Majorana's notation.) Following Majorana, the equation for $f(\tau )$, Eq.~(%
	\ref{MajoranaDiffEq1}), can be solved by the two-sided Laplace transform, 
	\begin{equation}
		\mathcal{L}[f(\tau )]=\int_{-\infty }^{\infty }\!\!\!e^{-s\tau }f(\tau )\;d\tau
		=F(s).
	\end{equation}%
	Here $F(s)$ is the Laplace transform of the function $f(\tau )$. Then we
	substitute this in Eq.~(\ref{MajoranaDiffEq1}), use the theorem about
	differentiation of the original function, \newline$\mathcal{L}[\tau f(\tau
	)]=-F^{\prime }(s)$, and obtain 
	\begin{equation}
		s^{2}F(s)-2i\left[ F(s)+sF^{\prime }(s)\right] +2\delta F(s)=0.
	\end{equation}%
	The solution of this first-order differential equation gives 
	\begin{equation}
		F(s)=C_{\delta }\exp{\left(-\frac{is^{2}}{4}\right)}s^{-1-i\delta }.
	\end{equation}%
	Here the constant of integration $C_{\delta }$ could be defined from an
	initial condition. And then we can find $f(\tau )$ from the inverse Laplace
	transform: 
	\begin{equation}
		f(\tau )=\lim_{T\rightarrow \infty }\int_{\gamma -iT}^{\gamma +iT}e^{s\tau
		}F(s)\;ds,  \label{f}
	\end{equation}%
	where $\gamma $ is a real number so that the contour path of integration is
	in the region of convergence of $F(s)$. This integral can be calculated by
	the steepest descent method \cite{Fedoruk1977}. For the following
	calculations, the contour could be deformed due to the residue theorem. This
	requires large times and we need to find the solution in two limits: for
	large positive time, which means that $\tau \rightarrow +\infty $, and for
	large negative time $\tau \ll 0$. Then, the integration contour in Eq.~(\ref%
	{f}) is over either the contour $L_{1}$ for $\tau <0$ or $L_{2}$ for $\tau
	>0 $, the contours to be defined. How the contours are chosen is described
	in detail in Appendix~\ref{AppendixA}; these steepest-descent contours $%
	L_{1,2}$ are demonstrated in Fig.~\ref{Fig:Majorana contour}.
	
	Our integral \eqref{f} has two contributions: the first one is from the saddle point and the second one is from the vicinity of zero. Details of the calculations are presented in Appendix~\ref{AppendixA}.
	
	First, we can write the contribution from the saddle point for $f(\tau )$
	as follows
	
	\begin{equation}
		f(\tau )=C_{\delta }\sqrt{{4\pi }}(-2i\tau )^{-1-i\delta }\exp\!{\left(i\frac{3\pi }{4}-i\tau ^{2}\right)}.
	\end{equation}%
	
	Second, we describe the contribution from the near-zero region $%
	s\rightarrow 0$, which is the integral Eq.~(\ref{f}) on the contour $L_{0}$. Here $%
	L_{0}$ is the near-zero vicinity contour with the neckline along the
	negative axis which is a part of the $L_{2}$ contour. Here we can neglect
	the term $s^{2}/4$ next to $s\tau $, and then we use the Gamma function in
	the Hankel integral representation \cite[\S 12.22]{Whittaker1920}: 
	\begin{equation}
		\int_{L_{0}}e^{x}x^{-y}dx\approx \frac{2\pi i}{\Gamma (y)}.
	\end{equation}%
	For using this, we define the replacement $x=s\tau $ and then for the
	integral in Eq.~(\ref{f}) we have 
	\begin{equation}
		\int_{L_{0}}\tau ^{i\delta }\frac{e^{x}}{x^{1+i\delta }}dx=\tau ^{i\delta }%
		\frac{2\pi i}{\Gamma (1+i\delta )}.
	\end{equation}
	
	So, we obtain the approximate solution of Eq.~\eqref{MajoranaDiffEq1} for
	two cases, $\tau <0$ and $\tau >0$, in the general form. The second part of
	the spinor, $g(\tau )$, is obtained from Eq.~(\ref{initialsystem})
	neglecting the terms $\sim \tau ^{-2}$ because this result is asymptotic.
	Then using the substitutions Eq.~\eqref{MajSubst}, we obtain 
	\widetext
	\begin{subequations}
		\label{evolution}
		\begin{align}
			& \tau <0:%
			\begin{cases}
				\alpha (\tau )=C_{\delta }\sqrt{4\pi }(-2i\tau )^{-i\delta -1}\exp\!\left( -i%
				\frac{\tau ^{2}}{2}+i\frac{3\pi }{4}\right) , \\ 
				\beta (\tau )=C_{\delta }\sqrt{\frac{2\pi }{\delta }}(-2i\tau )^{-i\delta
				}\exp\!\left( -\frac{i\tau ^{2}}{2}+\frac{i\pi }{4}\right) ,%
			\end{cases}
			\label{a} \\
			& \tau >0:%
			\begin{cases}
				\alpha (\tau )=C_{\delta }\sqrt{4\pi }(-2i\tau )^{-i\delta -1}\exp\!\left( -%
				\frac{i\tau ^{2}}{2}+i\frac{3\pi }{4}\right) +C_{\delta }\frac{2\pi i}{%
					\Gamma (i\delta +1)}\tau ^{i\delta }\exp\!\left( \frac{i\tau ^{2}}{2}\right) ,
				\\ 
				\beta (\tau )=C_{\delta }\sqrt{\frac{2\pi }{\delta }}(-2i\tau )^{-i\delta
				}\exp\!\left( -\frac{i\tau ^{2}}{2}+\frac{i\pi }{4}\right) +C_{\delta }\sqrt{%
					\frac{\delta }{2}}\frac{2\pi i}{\Gamma (i\delta +1)}\tau ^{i\delta -1}\exp\!				\left( \frac{i\tau ^{2}}{2}\right) .%
			\end{cases}
			\label{b}
		\end{align}%
		\endwidetext
		
		We now consider the initial condition far from the avoided-level crossing, at $\tau
		\rightarrow -\infty $, and from Eq.~(\ref{a}) obtain 
	\end{subequations}
	\begin{equation}
		\begin{cases}
			|\alpha |^{2}=0 \\ 
			|\beta |^{2}=1%
		\end{cases}
		.  \label{initialcondition}
	\end{equation}
	To fulfil the normalization condition we have taken the constant of
	integration $C_{\delta }$ as
	
	\begin{equation}
		C_{\delta }=\sqrt{\frac{\delta }{2\pi }}e^{-\frac{\pi \delta }{2}},
	\end{equation}%
	where we used that $i^{i\delta }=e^{-\pi \delta /2}$. Note that the initial
	condition (\ref{initialcondition}) leaves the phase undefined, so that
	replacing $C_{\delta }\rightarrow C_{\delta }e^{i\vartheta }$ with any phase $%
	\vartheta ~$would result in the same initial condition.
	
	\section{Probability, phase, and adiabatic-impulse model}
	\label{PPAAIM}
	\subsection{Asymptotic solution}
	
	~Based on the general equations above, we consider the limiting case. Omitting the terms $\sim \tau ^{-1}$, from Eq.~(\ref{evolution}), we obtain the asymptotes for $\alpha (\tau )$ and $\beta (\tau
	)$
	
	\begin{align}
		& \alpha (\tau \rightarrow -\infty )\rightarrow 0,
		\label{MajApproxSollution} \\
		& \beta (\tau \rightarrow -\infty )\rightarrow \left( -2i\tau \right)
		^{-i\delta }\exp\!\left( \frac{i\pi }{4}-\frac{\pi \delta }{2}-\frac{i\tau
			^{2}}{2}\right) ,  \notag \\
		& \alpha (\tau \rightarrow \infty )\rightarrow \frac{\sqrt{2\pi \delta }}{%
			\Gamma (1+i\delta )}\tau ^{i\delta }\exp\!{\left( -\frac{\pi \delta }{2}+%
			\frac{i\pi }{2}+\frac{i\tau ^{2}}{2}\right) },  \notag \\
		& \beta (\tau \rightarrow \infty )\rightarrow \left( -2i\tau \right)
		^{-i\delta }\exp\!\left( \frac{i\pi }{4}-\frac{\pi \delta }{2}-\frac{i\tau
			^{2}}{2}\right) .  \notag
	\end{align}%
	The expressions for $\beta (\tau )$ before and after the transition are
	similar. It is important to note that according to the sign of $\tau $ these
	formulas have different absolute values far from the transition region. When $%
	\tau <0$, we see that the expression ibecomes
	
	\begin{equation}
		\beta (\tau \rightarrow -\infty )\rightarrow \left( 2|\tau |\right)
		^{-i\delta }\exp\!\left( \frac{i\pi }{4}-\frac{i\tau ^{2}}{2}\right) ,
	\end{equation}%
	the absolute value of this is 1. However, when $\tau >0$ the result is 
	\begin{equation}
		\beta (\tau \rightarrow \infty )\rightarrow \left( 2|\tau |\right)
		^{-i\delta }\exp\!\left( -\pi \delta+\frac{i\pi }{4}-\frac{i\tau ^{2}}{2}%
		\right) ,
	\end{equation}%
	then the absolute value of this expression is different from~$1$. The
	respective transition probability is 
	\begin{equation}
		\mathcal{P}=\left\vert \beta (\tau \rightarrow \infty )\right\vert
		^{2}=\exp\left[-2\pi \delta \right].  \label{LZSM}
	\end{equation}%
	This is known as the LZSM formula. In view that many (if not most) authors
	in this context refer to this as the LZ formula, we emphasize that Majorana
	published this very result in Ref.~\cite{Majorana1932} \emph{before} LZS~ 
	\cite{Landau1932a, Landau1932b, Zener1932, Stueckelberg1932}.
	
	Being interested not only in the transition probability but rather in
	finding the full wave function, including the phase, we rewrite Eq.~(\ref%
	{MajApproxSollution}) in the exponential form: 
	\begin{eqnarray}
		\alpha (\tau \rightarrow \infty )\!\! &\approx &\!\!\sqrt{1-\mathcal{P}}\exp %
		\left[ i\mathrm{Arg}\left[\Gamma (1-i\delta )\right]+\frac{i\tau ^{2}}{2}+i\delta \ln {%
			\tau }\right] \!\!,  \notag \\
		\beta (\tau \rightarrow \infty )\! &\approx &\!\sqrt{\mathcal{P}}\exp \left[ 
		\frac{i\pi }{4}-\frac{i\tau ^{2}}{2}-i\delta \ln {2\tau }\right] \!.
		\label{alfa}
	\end{eqnarray}%
	Interestingly, in \cite{Majorana1932}, Majorana obtained only the correct probability, Eq.~(\ref{LZSM}). The phase in Eq.~(\ref{alfa}) can be
	obtained from Majorana's formulas if we note that there is a typo in the
	result for function $f(\tau \rightarrow \infty )$ in Ref.~\cite{Majorana1932}%
	: we should correct the typo by replacing 
	\begin{equation}
		e^{-k/4i}\rightarrow \tau_{\mathrm{M}} ^{-k/4i},
	\end{equation}%
	and we have also to replace $\sqrt{k}\rightarrow-2 \sqrt{\delta}$ and $\tau _{\mathrm{M}}\rightarrow \tau /\sqrt{2}$ to obtain our Eq.~(\ref{alfa}).
	\subsection{Adiabatic-impulse model}
	Let us further extend our results above by introducing the
	\textit{adiabatic-impulse model} \cite{Damski2005,Damski2006, Shevchenko2010}. In brief, this model consists of adiabatic evolution far from the avoided-level crossing,
	described by the propagators $U_{\mathrm{ad}}$ and the impulse-type
	probabilistic transition in the point of the avoided-level crossing, see
	Fig.~\ref{Fig:Adiabatic Impulse Model}. The latter is described by the
	matrix $N$, and we will demonstrate now how to obtain this. Far from the
	quasicrossing point, the evolution could be described in the following way 
	\begin{equation}
		\left\vert\psi _{\mathrm{f}}\right\rangle=U_{\mathrm{ad}}(0,t_{\mathrm{f}})NU_{\mathrm{ad}}(t_{\mathrm{i}},0)\left\vert\psi _{\mathrm{i}}\right\rangle,
	\end{equation}%
	where $\psi _{\mathrm{i}}$ is the initial wave function with components given by Eq.~(\ref{MajApproxSollution}) when $-\tau \rightarrow \infty $, and 
	$\psi _{\mathrm{f}}$ is the final state with components given by Eq.~(\ref%
	{MajApproxSollution}) when $\tau _{\mathrm{M}}\rightarrow \infty $. The adiabatic evolution is described in Appendix~\ref{AppendixB}.
	\subsection{Non-adiabatic transition}	
	A \textit{non-adiabatic transition }is described by the transfer matrix, which is associated with a scattering matrix \cite{Moskalets2011} in scattering theory. (Note the analogy with Mach-Zehnder interferometer \cite{Oliver2005,Sillanpaeae2006,Burkard2010}.) The components of the transfer matrix are related to the amplitudes of the respective states of the system in energy space.  The diagonal elements correspond to the square root of the reflection coefficient $R$, and the off-diagonal elements correspond to the square root of the transmission
	coefficient $T$ and its complex conjugate: 
	\begin{equation}
		N=\left( 
		\begin{array}{cc}
			\sqrt{R} & \sqrt{T} \\ 
			-\sqrt{T}^{\ast } & \sqrt{R}%
		\end{array}%
		\right) .
	\end{equation}%
	%
	%
	%
	%
	In our problem we obtain for the diagonal from Eqs.~(\ref{MajApproxSollution}) elements 
	\begin{equation}
		R=\mathcal{P}\text{ \ \ and \ \ }T=(1-\mathcal{P})\exp \left( i2\varphi _{%
			\mathrm{S}}\right) ,
	\end{equation}%
	where $\varphi _{\mathrm{S}}$ is the Stokes phase 
	\begin{equation}
		\varphi _{\mathrm{S}}=\frac{\pi }{4}+\mathrm{Arg} \left[ \Gamma \left( 1-i\delta
		\right) \right] +\delta \left( \ln {\delta }-1\right) .
	\end{equation}
	
	To conclude this section, an avoided-level crossing is described by the
	adiabatic-impulse model. With the matrices $U$ and $N$, it is
	straightforward to generalize the model to multi-level systems, e.g. Ref.~%
	\cite{Suzuki2022}. This demonstrates the thesis of Ref.~\cite{Wilczek2014}
	that \textquotedblleft \textit{Majorana's method is very well adapted to
		generalizations involving multiple level crossings}.\textquotedblright

	\begin{figure*}[t]
		\centering{\includegraphics[width=2.05\columnwidth]{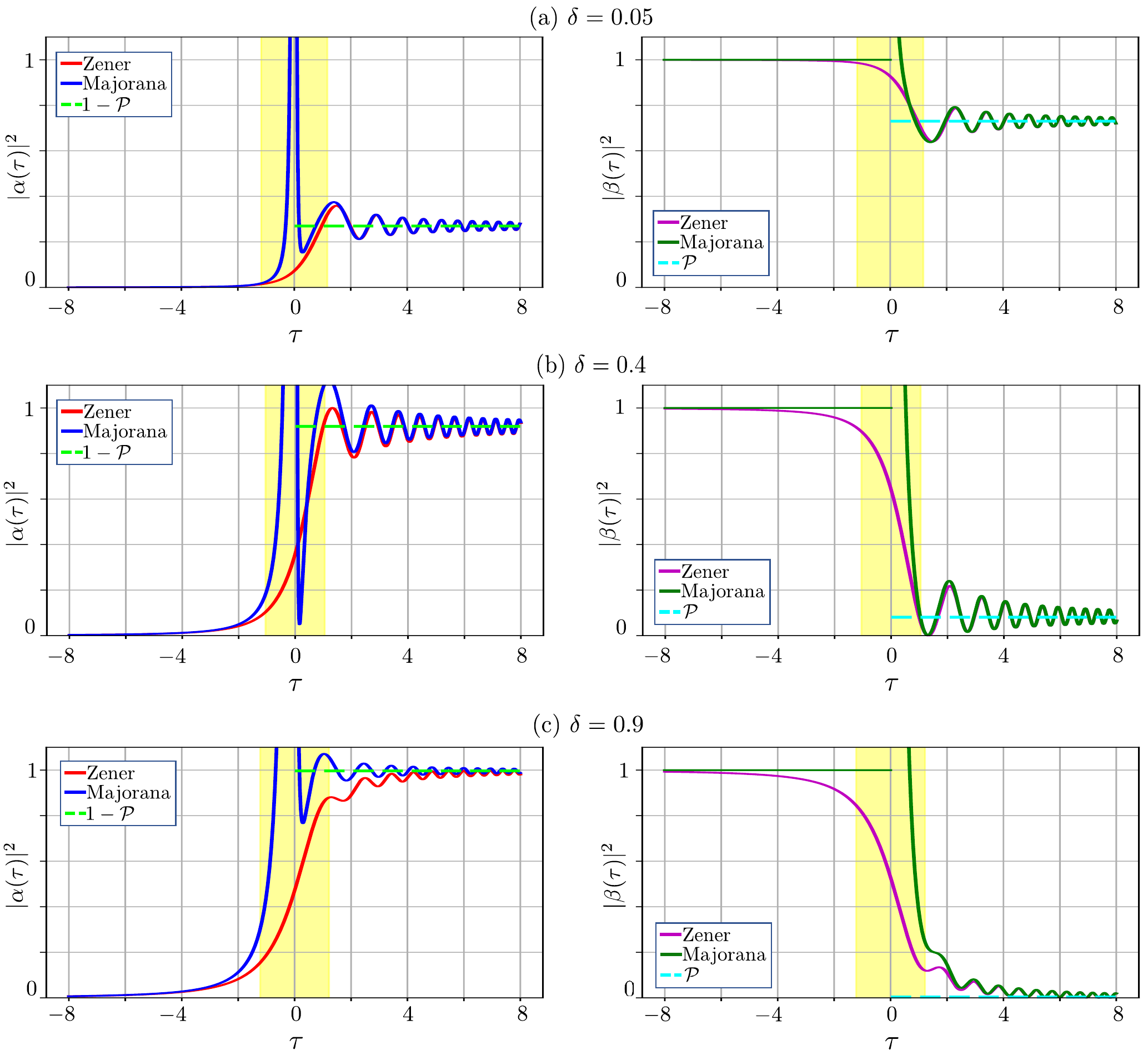}}
		\caption{Comparison of the dynamics of the occupation probability versus time $\protect\tau $, obtained by
			Majorana's method and Zener's method. The left panels show the dynamics of the
			first component of the spinor $\protect\alpha (\protect\tau )$, while the right panels
			show the second component $\protect\beta (\protect\tau )$, for three
			different values of the adiabaticity parameter $\protect\delta $. The yellow
			regions show the area where Majorana's approach does not give the correct result because that method is asymptotic. The bright green dashed lines show the LZSM probability. These illustrate that far from the
			transition region both results, by Zener and Majorana, tend to that value.}
		\label{Fig:dynamics}
	\end{figure*}
	
	\section{Comparison with Zener's approach}
	
	\label{CWZA} 
	As we wrote in the Introduction, out of the four approaches by LZSM~\cite{Majorana1932, Landau1932a, Landau1932b, Zener1932, Stueckelberg1932}, the total wave function is given only by the approaches by Zener and
	Majorana. The former is well known, while the latter is examined and extended here. Let us
	now compare the results by these two approaches.
	
	For readers' convenience, we write down here the final formulas of the
	Zener's approach (for details, see Ref.~\cite{Ivakhnenko2022} and references
	therein): 
	\begin{eqnarray}
		\alpha &=&A_{+}D_{-1-i\delta }\left( z\right) +A_{-}D_{-1-i\delta }\left(
		-z\right) ,  \label{ZenerAnsatz} \\
		\beta &=&B_{+}D_{-i\delta }\left( z\right) +B_{-}D_{-i\delta }\left(
		-z\right) .  \notag  \label{ZenExactSolving}
	\end{eqnarray}%
	Here $z=\tau \sqrt{2}e^{i\pi /4}$, $D_{\nu }(z)$ is the parabolic cylinder
	function, $B_{\pm }=\mp \delta ^{-1/2}\exp {\left( -i\pi /4\right) }A_{\pm }$%
	, and the coefficients $A_{\pm }$ are defined from an initial condition. We
	aim to compare the results obtained within Zener's approach with the ones obtained here employing Majorana's approach, Eq.~(\ref{evolution}). For this we	need to use the asymptotic behavior of a full analytical solution, see
	Appendix~\ref{AppendixC}, and apply the initial condition Eq.~(\ref{initialcondition}). As an impressive result, the asymptotic expressions for the wave function by Zener, \emph{coincide} with the ones we derived in Eq.~(\ref{evolution}) extending Majorana's approach.

	\begin{figure*}[t]
		\centering{\includegraphics[width=2.05 \columnwidth]{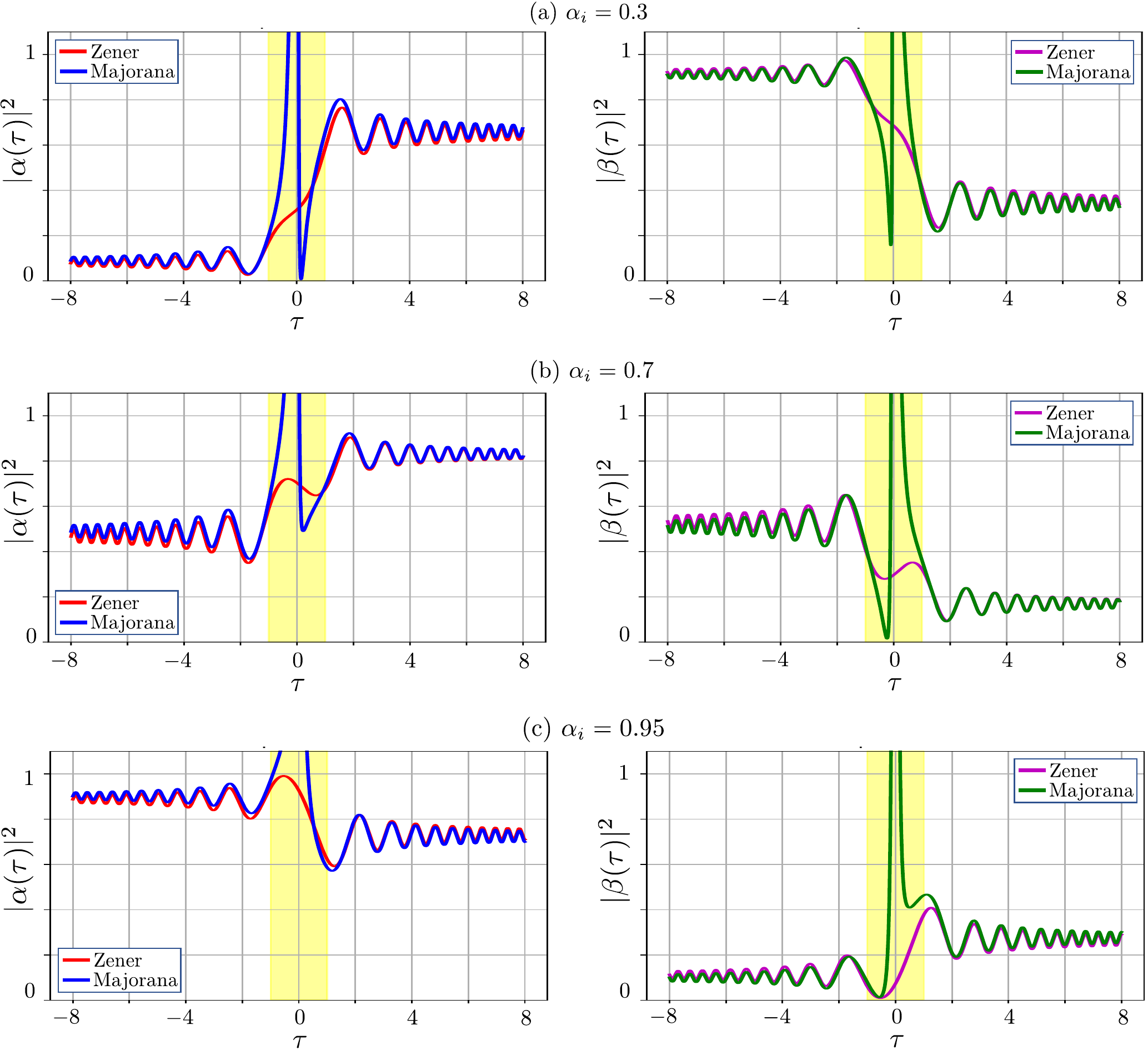}}
		\caption{Time-evolution of the occupation probability when starting from different initial superposition states. Similarly
			to Fig.~\protect\ref{Fig:dynamics}, we present the comparison of the dynamics obtained by following Majorana's method and Zener's method, now for three different
			initial states, with $\protect\alpha _{\mathrm{i}}=0.3$, $0.7$, and $0.95$; $%
			\protect\beta _{\mathrm{i}}=\protect\sqrt{1-\protect\alpha _{ \mathrm{i}}^{2}%
			}$. The adiabaticity parameter here is $\delta=0.1$.}
		\label{Fig:NonzeroInitialState}
	\end{figure*}
	
	\section{Dynamics}
	
	\label{Dynamics} For describing the dynamics of quantum systems, it is necessary to know the behaviour of the wave function for all times. Majorana obtained only the probability of the transition from the ground state to the excited one. We expand Majorana's method and obtain the dynamics of
	the wave function. Still, this method is asymptotic, and it is expected not to work appropriately at small absolute values of the dimensionless time $\tau $. Does this give the correct behaviour at finite values of time? Let us explore this and consider this dynamics, by plotting the energy-level occupations, given by Eqs.~(\ref{evolution}), as functions of time.
	
	In Fig.~\ref{Fig:dynamics} we show the occupations of the two levels
	for the asymptotic result of Majorana's approach, Eq.~\!(\ref{evolution}), and the exact result of Zener's approach, Eq.~\!(\ref{ZenerAnsatz}). As a nice surprise, Fig.~\ref{Fig:dynamics} shows that the asymptotic solution correctly describes the	dynamics at $\left\vert \tau \right\vert \gtrsim 1$; and this means that, even
	for relatively small times, Majorana's approach gives correct results. Thus, Majorana's approach \emph{not only describes correctly the asymptotic values at
		infinity} (which was the subject of Sec.~III), \emph{but also the transient dynamics.} Only in the very vicinity of zero, when $\left\vert \tau \right\vert
	\lesssim 1$, the asymptotic solution does not work correctly. As shown in  Fig.~\ref{Fig:dynamics}, this region, shown by the yellow background colour, is rather narrow.

	Let us quantify the region where our Majorana-type solution, Eq.~(\ref{evolution}), significantly deviates from the exact solution. As we can see
	from Fig.~\ref{Fig:dynamics}, this time interval corresponds to the
	so-called jump time \cite{Vitanov1999b, Ivakhnenko2022}. The jump time could
	be defined via the derivative at zero time, $P^{\prime }(0)$, in the
	following way for the Zener's approach 
	\begin{equation}
		\label{taujump}
		\tau _{\mathrm{jump}}=\frac{1-\mathcal{P}}{P^{\prime }(0)},
	\end{equation}%
	where $P(\tau )$ is the time-dependent transition probability from the lower
	level to the upper one. We obtain this probability as $P(\tau )=|\alpha
	(\tau )|^{2}$ from Eq.~(\ref{ZenerAnsatz}) using the initial condition Eq.~(%
	\ref{initialcondition}) 
	\begin{equation}
		P(\tau )=\delta \exp{\left(-\frac{\pi \delta }{4}\right)}\left\vert D_{-i\delta-1}(-z)\right\vert ^{2}.
	\end{equation}%
	Then the jump time depends on the adiabaticity parameter $\delta $ as
	follows 
	\begin{equation}
		\tau _{\mathrm{jump}}(\delta )=\frac{\sqrt{1-\mathcal{P}}}{\sqrt{2\delta }%
			\cos \chi (\delta )},  \label{tau_jump}
	\end{equation}%
	where 
	\begin{equation}
		\chi (\delta )=\frac{\pi }{4}+\mathrm{Arg}\left[ \Gamma \left( \frac{1}{2}-\frac{i\delta }{%
			2}\right) \right]-\mathrm{Arg}\left[ \Gamma \left( 1-\frac{i\delta }{2}\right) \right].
	\end{equation}%
	Based on this formula, we can analytically estimate the area where our
	method does not give the correct result: the width of the yellow-background area in Fig.~\ref{Fig:dynamics} corresponds to $\tau _{\mathrm{jump}}$, given by Eq.~(\ref{tau_jump}).
	
	\section{Arbitrary initial state}
	
	\label{AIS} When in Sec.~\ref{DAILT} we solved the evolution equations
	following Majorana, we obtained the specific initial condition, Eq.~(\ref%
	{initialcondition}). If one is interested in any other initial condition,
	this solution is invalid. This is because this approach gives a partial
	solution of Eq.~(\ref{MajoranaDiffEq1}). In order to find the general
	solution, we need to obtain a second partial solution, linearly independent
	from the first one. This we can obtain by solving Eq.~(\ref{MajoranaDiffEq2}%
	) analogously to how we did in Sec.~\ref{DAILT}. Let us call these two
	solutions $\left\vert \psi _{1}\right\rangle$ and $\left\vert \psi _{2}\right\rangle$ with 
	\begin{equation}
		\left\vert \psi _{1,2}\right\rangle=%
		\begin{pmatrix}
			\alpha _{1,2} \\ 
			\beta _{1,2}%
		\end{pmatrix}%
		.
	\end{equation}

	Instead of solving Eq.~(\ref{MajoranaDiffEq2}), we note that this coincides
	with Eq.~(\ref{MajoranaDiffEq1}) by swapping $\alpha _{2}$ with $\beta _{1}$
	and $\beta _{2}$ with $\alpha _{1}$ and taking their complex conjugate, with the
	appropriate choice of the branch in the square root. In the
	method of steepest descent, the branch depends on the inclination angle of the integration contour. In the two partial solutions, the inclinations are
	different, so the branches are also chosen to be different.
	
	Then the general solution is a linear combination of these two solutions, $\left\vert \psi (\tau )\right\rangle=Q_{1}\left\vert \psi_{1} (\tau )\right\rangle+Q_{2}\left\vert \psi_{2} (\tau )\right\rangle$. If we consider a
	given initial state 
	\begin{equation}
		\left\vert\psi (\tau _{\mathrm{i}})\right\rangle= 
		\begin{pmatrix}
			\alpha _{_{\mathrm{i}}} \\ 
			\beta _{_{\mathrm{i}}}%
		\end{pmatrix}
		,
	\end{equation}
	then the constants are $Q_{1}=\beta _{\mathrm{i}}$ and $Q_{2}=\alpha _{ 
		\mathrm{i}}$.

	To summarize, the general solution is 
	\begin{equation}
		\left\vert\psi (\tau )\right\rangle=\beta _{\mathrm{i}}\left\vert\psi _{1}(\tau )\right\rangle+\alpha _{\mathrm{i}}\left\vert\psi
		_{2}(\tau )\right\rangle,
		\label{GeneralSolution}
	\end{equation}
	with $\left\vert\psi _{1}(\tau )\right\rangle$ given by Eq.~(\ref{evolution}) and $\left\vert\psi _{2}(\tau )\right\rangle$
	obtained from $\left\vert\psi _{1}(\tau )\right\rangle$ by swapping $\alpha $ and $\beta $ and a
	subsequent complex conjugation.
	
	Results of calculations for \textit{superposition} initial states are shown in Fig.~%
	\ref{Fig:NonzeroInitialState}. There we consider three initial states, when the
	probability of the $\psi _{1}$-state, which is $\left\vert \alpha _{\mathrm{i}%
	}\right\vert ^{2}$, is close to $0.1$ (top panel), $0.5$ (middle panel), and 
	$0.9$ (bottom panel). The top panel describes the slight deviation from
	Majorana's solution $\left\vert\psi _{1}\right\rangle$, presented in Fig.~\ref{Fig:dynamics}, while
	the bottom panel describes the opposite case (which can be called anti-Majorana solution), when the solution is closer to $\left\vert\psi _{2}\right\rangle$.
	
\begin{figure*}[t]
		\centering{\includegraphics[width=2.1 \columnwidth]{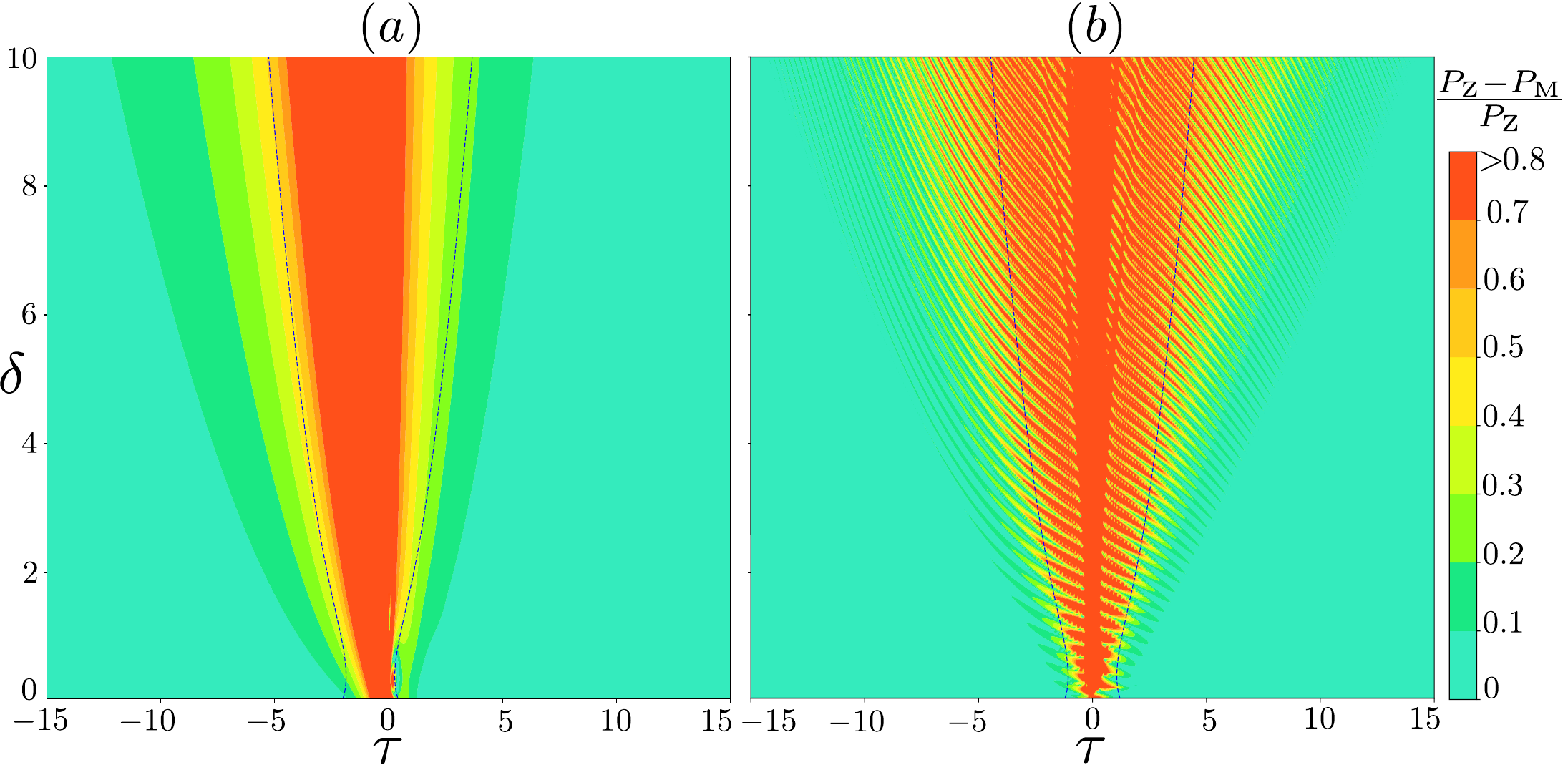}}
		\caption{Validity of Majorana's approach. This is quantified as the difference of the upper-level occupation probability calculated by both the asymptotic Majorana's method and the exact Zener's result. The initial condition is taken to be the ground state with $\protect\alpha _{\mathrm{i}}=0$ (a) and the superposition state (b) with $\protect\alpha _{\mathrm{i}}=0.7$. Here $P_\mathrm{Z}$ is Zener's probability which is given by $|\alpha(\tau)|^2$ from Eq.~(\ref{ZenerAnsatz}). Then $P_\mathrm{M}$ is our result within Majorana's approach which is provided by $|\alpha(\tau)|^2$ from Eq.~(\ref{GeneralSolution}). The dashed blue lines show the jump time $\tau_{\mathrm{jump}}$, Eq.~(\ref{taujump}). In general, the jump time is not symmetric around the point $\tau=0$; on the left panel the jump time is shifted to the left, for illustrative purposes.}
		\label{Fig:deviation}
	\end{figure*}

	Figure~\ref{Fig:deviation} shows the validity of our results obtained for a ground (a) and superposition state (b) within Majorana's approach. This is quantified as the relative difference between our asymptotic solution and the exact one obtained within Zener's approach. For illustrative purposes we choose the equally populated superposition at the initial time. If $\left\vert \alpha _{\mathrm{i}%
	}\right\vert ^{2}<0.5$, the figure becomes less symmetric and skewed to the left; if $\left\vert \alpha _{\mathrm{i}}\right\vert ^{2}>0.5$ then skewed to the right. In addition, we plot dashed blue lines, showing the jump time from Eq.~(\ref{taujump}), which gives a good estimate of where the asymptotic solution is in agreement with the exact one (outside of the region between the blue curves). Note that the jump-time is not symmetric around the point $\tau=0$. In general, it is asymmetric, and shift relates to the value of the adiabaticity parameter $\delta$.

	\section{Conclusions}
	
	We studied the approach by Majorana to the problem of a transition through a region where the energy levels experience avoided-level crossing. We demonstrated that \textquotedblleft \textit{Majorana's method is smooth and capable of considerable generalization}\textquotedblright\ \cite{Wilczek2014}, and we considered this here in detail. We obtained in a mathematically elegant way, the asymptotic formulas for the
	amplitude and phase of the two-level system wave function after a single
	passage with linear excitation. This was done using
	Laplace transformations and contour integration. Our study demonstrates that
	Majorana's approach justifies the adiabatic-impulse model. We demonstrated
	that this asymptotic model provides a very good description of the dynamics.
	
	\begin{acknowledgments}
		We gratefully acknowledge discussions with S.~Esposito, M.~Moskalets, V.~Riabov. 
		
		The research work of P.O.K., O.V.I., and S.N.S. is sponsored by the Army Research
		Office under Grant No.~W911NF-20-1-0261. P.O.K. and
		O.V.I. gratefully acknowledge IPA RIKEN scholarships. F.N. is supported in
		part by: Nippon Telegraph and Telephone Corporation (NTT) Research, the
		Japan Science and Technology Agency (JST) [via the Quantum Leap Flagship
		Program (Q-LEAP), and the Moonshot R\&D Grant Number JPMJMS2061], the Japan
		Society for the Promotion of Science (JSPS) [via the Grants-in-Aid for
		Scientific Research (KAKENHI) Grant No. JP20H00134], the Army Research
		Office (ARO) (Grant No. W911NF-18-1-0358), the Asian Office of Aerospace
		Research and Development (AOARD) (via Grant No. FA2386-20-1-4069), and the
		Foundational Questions Institute Fund (FQXi) via Grant No. FQXi-IAF19-06.
	\end{acknowledgments}
	
	\appendix
	
	\section{Integration contours}
	
	\label{AppendixA}
	
	We can use the saddle-point method for obtaining the result of the inverse
	Laplace transform, because our integrals are Laplace-type integrals \cite{Fedoruk1977}
	
	\begin{equation}
		I(\tau)=\int\limits_{L} \phi(z)\; e^{\tau \mu(z)} dz.
		\label{LaplaceInt}
	\end{equation}
	
	First, we must define the integration contours. These should be situated in
	the holomorphic region of the sub-integral function. The function $F(s)$ contains an exponential factor, which means that we should define the holomorphy region for a logarithm. This region is the whole complex plane with a cut along the real axis from
	minus infinity to zero.
	
	The contour $L$ should be a steepest-descent contour of an integral for
	applying the saddle-point method. For this, the
	following conditions should be satisfied~\cite{Fedoruk1977}:\newline
	1. There is a point $z_0$ belonging to the contour $L$ such that $\mbox{Re}\left[\mu (z_{0})\right]=\max\limits_{z\in L}\mbox{Re}\left[\mu
	(z)\right]$, where the function $\mu(\tau)$ was defined in the integral Eq.~(\ref{LaplaceInt});\newline
	2. There is such a value $\tau_0>0$ that the following integral has a finite value when integrated over the contour $L$ 
	\begin{equation}
		\int\limits_{L}|\phi (z)|\exp %
		\left[ \tau _{0}\mbox{Re}\mu (z)\right] dz<\infty ;
	\end{equation}
	3. For a function $\mu(\tau)$ the following conditions are satisfied: 
	\begin{eqnarray}
		\mu ^{\prime }(z_{0})&=&0, \\
		\notag
		\mu ^{\prime \prime }(z_{0})&\neq&0,  \\
		\frac{d^{2}}{dy^{2}}\left[ \mbox{Re}\mu (z_{0}+y\lambda )\right]&=&\mathrm{Const}<0,
		\notag  
	\end{eqnarray}
	where $\lambda $ is any tangent to contour $L$ in the point $z_{0}$;\newline
	4. Functions $\phi $ and $\mu $ are holomorphic in the vicinity of the
	contour.
	
	The saddle-point method implies that if the contour $L$ is a steepest-descent
	contour for $I(\tau )$, then 
	\begin{equation}
		I(\tau )=\sqrt{\frac{2\pi }{\tau }}\frac{\phi (z_{0})e^{\tau \mu (z_{0})}}{%
			\sqrt{-\mu ^{\prime \prime }(z_{0})}}\left( 1+o(1)\right) .
	\end{equation}%
	A branch of a square root $\sqrt{-\mu ^{\prime \prime }(z_{0})}$ should be chosen
	corresponding to the angle of inclination of the integration contour. We
	should make a substitution for working with the steepest-descent contour in
	our problem 
	\begin{equation}
		s=\tau z,\text{ \ \ }ds=\tau dz;
	\end{equation}%
	then 
	\begin{equation}
		f(\tau )=C_{\delta }\tau ^{-i\delta }\int\limits_{L}\exp \left\{ \tau
		^{2}\left( z-i\frac{z^{2}}{4}\right) \right\} z^{-(i\delta +1)}dz,
	\end{equation}%
	\begin{equation}
		\phi (z)=z^{-(i\delta +1)},
	\end{equation}%
	\begin{equation}
		\mu (z)=z-i\frac{z^{2}}{4}.
	\end{equation}
	
	From the condition \#$3$ listed above, we obtain the point $z_0$
	
	\begin{equation}
		z_{0}=-2i\;\;\Rightarrow\;\; s_{0}=-2i\tau .
	\end{equation}%
	Also, we can obtain the angle of inclination of the tangent from the second
	part of the condition \#$3$:%
	\begin{equation}
		\mathrm{Re}\left( -\lambda ^{2}\frac{i}{2}\right) <0\;\;\Rightarrow \;\;\mathrm{Arg} \lambda
		=\frac{3\pi }{4}.
	\end{equation}%
	
	From the condition \#$1$ above we obtain the regions where the saddle-point method could be
	applied, defined by
	\begin{equation}
		\mathrm{Re}\left[\mu (z_{0})\right]-\mathrm{Re}\left[\mu (z)\right]>0.
	\end{equation}%
	The regions corresponding to this condition are highlighted by the light-green	colour background in Fig.~\ref{Fig:Majorana contour}. We need to consider two different contours: with $\tau >0$ and $\tau <0$. In case of $\tau >0$, there is a part of the contour which lays outside of the green area. According to it,
	in this case the result comes not only from the saddle-point method but also
	from the integration near the zero point. 
	
	\section{Adiabatic evolution}
	
	\label{AppendixB} In this section we discuss the adiabatic evolution in the
	diabatic basis. This is needed for the adiabatic-impulse model, which we
	formulate and justify in Sec.~\ref{PPAAIM}. Here, the adiabatic evolution
	consists in staying in one of the adiabatic eigenstates $\left\vert\varphi _{\pm }(t)\right\rangle$%
	, which are defined as eigenstates of $H(t)$, 
	\begin{equation}
		H(t)\left\vert\varphi _{\pm }(t)\right\rangle=E_{\pm }(t)\left\vert\varphi _{\pm }(t)\right\rangle.
	\end{equation}%
	Solving the non-stationary Schr\"{o}dinger equation, $i\hbar \dot{\left\vert{\psi}\right\rangle}=E\left\vert\psi\right\rangle$, we obtain \cite{Shevchenko2010} 
	\begin{equation}
		\left\vert\varphi _{\pm }(t)\right\rangle=\left\vert\varphi _{\pm }(t_\mathrm{i})\right\rangle\exp {\left\{ \mp i\left( \zeta +%
			\frac{\pi }{4}\right) \right\} },  \label{adisbsticWF}
	\end{equation}%
	\begin{equation}
		\zeta =\frac{1}{2\hbar }\int_{t_\mathrm{i}}^{t}{\Delta E(t^{\prime })dt^{\prime }},
	\end{equation}%
	\begin{equation}
		\Delta E(t)=E_{+}(t)-E_{-}(t)=\sqrt{\Delta ^{2}+\varepsilon ^{2}(t)}.
	\end{equation}%
	The full wave function is 
	\begin{equation}
		\left\vert\psi (t)\right\rangle=b_{+}(t)\left\vert\varphi _{+}(t_\mathrm{i})\right\rangle+b_{-}(t)\left\vert\varphi _{-}(t_\mathrm{i})\right\rangle,
	\end{equation}%
	where $b_{\pm }$ are the respective amplitudes. In the adiabatic basis we
	can write the wave function as a vector 
	\begin{equation}
		\left\vert\psi(t)\right\rangle =\left( 
		\begin{array}{c}
			b_{+}(t) \\ 
			b_{-}(t)%
		\end{array}%
		\right) .  \label{psi}
	\end{equation}%
	This allows us to describe the adiabatic evolution from the time moment $t_\mathrm{i}$ to $t_\mathrm{f}$ with the evolution matrix $U_\mathrm{ad}$:
	\begin{equation}
		\left( 
		\begin{array}{c}
			b_{+}(t_{\mathrm{f}}) \\ 
			b_{-}(t_{\mathrm{f}})%
		\end{array}%
		\right) =U_{\mathrm{ad}}\left( 
		\begin{array}{c}
			b_{+}(t_{\mathrm{i}}) \\ 
			b_{-}(t_{\mathrm{i}})%
		\end{array}%
		\right) .
	\end{equation}%
	The adiabatic evolution is then obtained from Eqs.~(\ref{psi}) and~(\ref%
	{adisbsticWF}):
	\begin{equation}
		U_{\mathrm{ad}}=\left( 
		\begin{array}{cc}
			\exp \left( -i\zeta \right) & 0 \\ 
			0 & \exp \left( i\zeta \right)%
		\end{array}%
		\right) .
	\end{equation}
	
	In our problem the bias is linear in time $\varepsilon (t)=vt$ and we can
	calculate the asymptotic expressions for $\zeta $ at large times, i.e. at $%
	t=\pm \tau _{\mathrm{a}}\sqrt{2\hbar /v}$, with $\tau _{\mathrm{a}}\gg 1$: 
	\begin{eqnarray}
		\zeta \left( \pm \tau _{\mathrm{a}}\right) &=&\frac{1}{2\hbar }%
		\int\limits_{0}^{\pm \tau _{\mathrm{a}}}\sqrt{\Delta ^{2}+\varepsilon ^{2}}%
		d\tau \approx \\
		&\approx &\pm \left[ \frac{\tau _{\mathrm{a}}^{2}}{2}+\frac{\delta }{2}-%
		\frac{\delta }{2}\ln {\delta }+\delta \ln {\sqrt{2}\tau _{\mathrm{a}}}\right]
		.
	\end{eqnarray}%
	The diabatic states $\left\vert\psi _{\pm }\right\rangle$ are the eigenstates of the Hamiltonian
	with $\Delta =0$, which means 
	\begin{equation}
		\sigma _{z}\left\vert\psi _{\pm }\right\rangle=\pm\left\vert \psi _{\pm }\right\rangle\text{.}
	\end{equation}%
	We work in the diabatic basis so we need to transfer the adiabatic evolution
	matrix from the adiabatic basis to the diabatic one. The relation between the bases
	is \cite{Shevchenko2010} 
	\begin{equation}
		\left\vert\varphi _{\pm }(t)\right\rangle=\gamma _{\mp }\left\vert\psi _{+}\right\rangle\mp \gamma _{\pm }\left\vert\psi _{-}\right\rangle,
	\end{equation}%
	where
	\begin{equation}
		\gamma _{\pm }=\frac{1}{\sqrt{2}}\sqrt{1\pm \frac{\varepsilon (t)}{\Delta
				E(t)}}.
	\end{equation}%
	This relation can be simplified far from the transition region, at $|\tau |\gg 1$. Then, the adiabatic evolution matrix in the diabatic basis has two different forms, before the transition,
	\begin{equation}
		\tau <0:\,\,\,\,U_{\mathrm{ad}}(\tau_\mathrm{i}, 0)=\left( 
		\begin{array}{cc}
			\exp \left( -i\zeta \right) & 0 \\ 
			0 & \exp \left( i\zeta \right)%
		\end{array}%
		\right)
	\end{equation}%
	and after it,
	\begin{equation}
		\tau >0:\,\,\,\,U_{\mathrm{ad}}(0, \tau_\mathrm{f})=\left( 
		\begin{array}{cc}
			\exp \left( i\zeta \right) & 0 \\ 
			0 & \exp \left( -i\zeta \right)%
		\end{array}%
		\right) .
	\end{equation}%
	
	Then the matrix for the overall evolution in the diabatic basis is described by the following matrix 
	\begin{equation}
		U_{\mathrm{ad}}(0,\tau_{\mathrm{f}})NU_{\mathrm{ad}}(\tau_\mathrm{i}, 0)=%
		\begin{pmatrix}
			\sqrt{R} & \sqrt{T}e^{2i\zeta (\tau _{a})} \\ 
			-\sqrt{T}^{\ast }e^{-2i\zeta (\tau _{a})} & \sqrt{R}%
		\end{pmatrix}%
		.
	\end{equation}
	
	\section{Asymptotics of Zener's wave function}
	
	\label{AppendixC} Zener's approach gives the full analytical solution in
	terms of the parabolic cylinder functions, e.g.~\cite{Ivakhnenko2022}, 
	\begin{equation}
		\begin{cases}
			\alpha =A_{+}D_{-i\delta -1}(z)+A_{-}D_{-i\delta -1}(-z), \\ 
			\beta =-\frac{A_{+}}{\sqrt{\delta }}\exp \!\left( -i\frac{\pi }{4}\right)
			\!D_{-i\delta }(z)+\frac{A_{-}}{\sqrt{\delta }}\exp \!\left( -i\frac{\pi }{4}%
			\right) \!D_{-i\delta }(-z),%
		\end{cases}
		\label{alpha_beta}
	\end{equation}%
	where $z=\tau \sqrt{2}e^{\frac{i\pi }{4}}$. The coefficients are obtained
	from an initial condition at $z=z_{\text{i}}$, 
	\begin{eqnarray}
		A_{+} &=&\frac{\Gamma (1+i\delta )}{\sqrt{2\pi }}\left[ \alpha (z_{\text{i}%
			\!\!}\ )D_{-i\delta }(-z_{\text{i}})-\right.  \notag \\
		&&\left. -\beta (z_{\text{i}\!\!}\ )e^{i\frac{\pi }{4}}\sqrt{\delta }%
		D_{-1-i\delta }(-z_{\text{i}})\right] , \\
		A_{-} &=&\frac{\Gamma (1+i\delta )}{\sqrt{2\pi }}\left[ \alpha (z_{\text{i}%
			\!\!}\ )D_{-i\delta }(z_{\text{i}})+\right.  \notag \\
		&&\left. +\beta (z_{\text{i}\!\!}\ )e^{i\frac{\pi }{4}}\sqrt{\delta }%
		D_{-1-i\delta }(z_{\text{i}})\right] .
	\end{eqnarray}
	
	Comparing the time evolution obtained in our work following Majorana's
	approach, Eq.~(\ref{evolution}), with Zener's wave function requires to take
	the asymptotic expressions of these formulas. The asymptotes of the
	parabolic cylinder function are as follows (from Ref.~\cite{GradshteynRyzhik}%
	), depending on $\mathrm{Arg} {z}$,
	
	1) for $-\frac{5\pi }{4}<\mathrm{Arg} {z}<-\frac{\pi }{4}$\newline
	\begin{equation}
		D_{p}(z)\sim \exp{\left(-\frac{z^{2}}{4}\right)}z^{p}-\frac{\sqrt{2\pi }}{\Gamma (-p)}
		e^{-ip\pi }\exp{\left(\frac{z^{2}}{4}\right)}z^{-p-1}  \label{asymptotic1}
	\end{equation}
	2) and for $|\mathrm{Arg} z|<\frac{3\pi }{4}$\newline
	\begin{equation}
		D_{p}(z)\sim \exp{\left(-\frac{z^{2}}{4}\right)}z^{p}.  \label{asymptotic2}
	\end{equation}
	
	There are two cases. For $\tau <0$, we have 
	\begin{equation}
		z=\left( -\left\vert \tau \right\vert \right) \sqrt{2}\exp{\left(\frac{i\pi }{4}
			\right)}=\left\vert \tau \right\vert \sqrt{2}\exp{\left(-\frac{i3\pi }{4}\right)}
	\end{equation}
	and then Eq.~(\ref{asymptotic1}) applies. For $\tau >0$ we have 
	\begin{equation}
		z=\tau \sqrt{2}\exp{\left(\frac{i\pi }{4}\right)},
	\end{equation}
	and then Eq.~(\ref{asymptotic2}) applies.
	
	Using these asymptotic expressions for Eq.~(\ref{alpha_beta}), we obtain
	exactly Eq.~(\ref{evolution}), which shows that the result using Majorana's approach provides the
	very same asymptotic time evolution as the one by Zener's approach.
	
	\newpage
	\nocite{apsrev41Control} 
	\bibliographystyle{apsrev4-1}
	\bibliography{LZSM2,1}
	
\end{document}